\documentclass[12pt]{article}
\usepackage{epsfig}
\usepackage{latexsym}
\begin{document}

\title{New approach to the quantization of the Yang-Mills field.}
\author{ A.A.Slavnov \thanks{E-mail:$~~$ slavnov@mi.ras.ru}\\
V.A.Steklov Mathematical Institute\\ Gubkina street 8, GSP-1,119991,\\
and Moscow State University}

\maketitle

\begin{abstract}
Review of the papers on the new method of the Yang-Mills field quantization applicable both in perturbation theory and beyond it is presented.  It is shown that in the modified formulation of the Yang-Mills theory leading to the formal perturbation theory, which coincides with the standard one, there exist soliton solutions of the classical equations of motion.
\end{abstract}

\section{Introduction.}

The progress in physics as a rule is related to the introduction of the new symmetry. Very small amount of problems of the high energy physics may be solved exactly. However on the basis of symmetry one can make some predictions, which can be checked experimentally. The good illustration of this thesis is given by the gauge field theory. Classical electrodynamics of Faradey-Maxwell is described completely by the electromagnetic stress tensor $F_{\mu\nu}$. However if we would like to describe the interaction of the electromagnetic field with the matter fields, for example with electron, we shall discover that using only the stress tensor $F_{\mu\nu}$, it is impossible to construct a local Hermitean Hamiltonian, which describes this interaction. For this reason electromagnetic field is described by the four vector $A_\mu $ , through which the stress tensor may be expressed. The vector $A_\mu $ has more components than needed for the description of experiment: it is known that electromagnetic field is three dimensionally transversal, that is the electric and magnetic polarization vectors are perpendicular to the three dimensional momentum. From the point of view of experiment it is necessary to introduce only three dimensionally transversal components of vector $A_\mu $. But the vector $A_\mu $ except for the transversal components has also the time-like componentïî  $A_0 $, and the component parallel to the three dimensional momentum, which usually is denoted as $A_3 $. Clearly the theory, based on the vector $A_\mu $ has more components than necessary. However simultaneously with the vector $A_\mu $ the new symmetry - gradient (gauge) invariance is introduced. This invariance provides the decoupling of the components $A_0 $ and $A_3 $ from the transversal components and observables do not depend on $A_0 $ è $A_3 $. The same idea may be illustrated by the models based on the nonabelian gauge groups. The simplest version of such a theory is based on the group $SU(2)$, and was proposed by Yang and Mills (\cite{YM}). After the transition to the quantum theory in the Yang-Mills model except for unphysical components of the vector field the anticommuting scalar fields $\bar{c}, c$ - Faddeev-Popov ghosts arise (\cite{FP}) , (\cite{DW}). One can show(\cite{KO}), that the components $A_0, A_3$ and Faddeev-Popov ghosts decouple from the transversal components and observables do not depend on unphysical particles. 

The same idea is the foundation of renormalizable description of the Higgs model (\cite{BE}, \cite{H}, \cite{K}), which allows without breaking the gauge invariance to introduce the mass term for the vector field. In this case the complete spectrum of the theory includes except for $A_i$ and Faddeev-Popov ghosts, also the Goldstone particles $B_a$. As before one can show that the component $A_0$, Faddeev-Popov ghosts  
$\bar{c}, c$ and Goldstone fields decouple from the physical components of the vector field $A_i, \quad i=1,2,3$  and observables depend only on these components.

In these examples we observe the following law: for consistent description of observables it is necessary to enlarge the spectrum of the theory introducing unphysical excitations. Of course we should care that the new symmetry arised in the theory, leading to decoupling of unphysical degrees of freedom from the physical ones.

We are going to use this observation to construct the scheme of quantization of nonabelian gauge fields, applicable beyond perturbation theory. We shall show also that the Yang-Mills theory in the modified formulation, leading in the quantum case to the same formal perturbation theory as the standard one, has soliton solutions of the classical equations of motion. It contradicts to the generally accepted point of view, according to which the classical Yang-Mills theory does not possess soliton solutions , and is in accordance with the hypothesis that the confinement of the color objects is related to the existence of the quasi particle solutions of classical equations which have the localized finite energy. 

Speaking about the quantization of gauge fields beyond perturbation theory, we have in mind the problem of non uniqueness of quantization, noticed for the first time by V.N.Gribov (\cite{Gr}). It is known that to quantize the gauge field it is necessary to impose some gauge condition choosing the unique representative in the class of gauge equivalent configurations. In the Maxwell theory that means that the gauge condition (for example the Coulomb gauge)
\begin{equation}
\partial_iA_i=0
\label{1}
\end{equation}
has only trivial solution for the function $\Phi$, where gauge equivalent configurations look as follows
\begin{equation}
A_i+\partial_i\Phi
\label{2}
\end{equation}
Indeed, if the condition (\ref{1}) is fulfilled, the function by which differ two gauge equivalent configurations, should satisfy the equation
\begin{equation}
\triangle\Phi=0
\label{3}
\end{equation}
that is to be the harmonic function. It is known that the harmonic function has extremal value at the border. As $\Phi(x)$ must vanish at the spatial infinity we conclude that this function is identically zero. Therefore in the case of Abelian gauge group the Coulomb gauge selects a unique representative in the class of gauge equivalent configurations. 

However in the case of the simplest nonabelian gauge group $SU(2)$ the equation, corresponding to (\ref{3}) looks as follows
\begin{equation} 
\triangle\Phi^a+g\partial_i (\varepsilon^{abc}A_i^b \Phi^c)=0
\label{4}
\end{equation}
This equation even if $\Phi(x) \rightarrow 0$ when $|x| \rightarrow \infty$ has nontrivial solutions. In the framework of perturbation theory with respect to the coupling constant the solution of the equation (\ref{4}) is trivial. Indeed, if we shall look for the solution for $\Phi$ as a formal series
\begin{equation}
\Phi=\Phi_0+g\Phi_1+g^2\Phi_2+\ldots
\label{5}
\end{equation}   
than in the framework of perturbation theory we shall get     
\begin{equation}
\Phi_0=\Phi_1=\Phi_2= \ldots=0
\label{6}
\end{equation}
For large $g$ the equation (\ref{4}) has nontrivial solutions, which tend to zero at spatial infinity. Therefore the quantization scheme of Faddeev-Popov-De Witt, based on the assumption, that the Coulomb gauge condition chooses a unique representative in the class of gauge equivalent configurations, strictly speaking is not applicable beyond the perturbation theory. If we shall try to use it for large g, this will lead to appearance of singularities in the path integral for the scattering matrix.

I.Singer generalized this result for any differential gauge (\cite{Si}).

One can hope that it is possible to use so called algebraic gauges, for example the Hamiltonian gauge $A_0=0$. In these cases the problem of uniqueness of the gauge choice  does not arise. But putting in the Lagrangian
 $A_0=0$, we loose the constraint $D_iP_i^a=0$, which should be fulfilled for the observable quantities. It is impossible to impose this condition on the fields, which are assumed in the process of quantization to be independent. One can impose this condition on the allowed state vectors
 \begin{equation}
\hat {D_i }\hat{P_i}|\Phi>=0 
\label{7}
\end{equation}
This condition can be solved in perturbation theory, but the question about the existence of solutions of (\ref{7}) beyond perturbation theory is open. Analogous problems arise in other algebraic gauges. Moreover, from the practical point of view these gauges are not satisfactory because they destroy the explicit Lorentz invariance which complicates calculations considerably. So we see that in the standard formulation the quantization of nonabelian gauge fields is possible only in the framework of perturbation theory. A possible way out of this situation was proposed in the series of papers by Zwanziger (\cite{Zw}).

In this paper we are going to use the other possibility. We shall show that due to the mechanism, described above, that is expanding the spectrum of unphysical fields and introducing the new symmetry one can quantize
nonabelian gauge theory beyond perturbation theory. The modified theory coincides with the standard one in the framework of formal perturbation theory, but has soliton solutions of the classical equation of motions of the type of t'Hooft-Polyakov magnetic monopole (\cite{Ho}, \cite{Po}).

\section{ The modified Yang-Mills theory.}

We start with the modified $SU(2)$ theory, described by the Lagrangian
 \begin{equation}
L=-\frac{1}{4}F^i_{\mu\nu}F^i_{\mu\nu}+(D_\mu \varphi^+)^i(D_\mu \varphi^-)^i+i(D_\mu b)^i(D_\mu e)^i
\label{8}
\end{equation}
Here $F^i_{\mu\nu}$ is the usual stress tensor for the Yang-Mills field, and $D_\mu$ denotes a covariant derivative. The fields $\varphi^\pm, b, e$ - Hermithean  and have zero spin. The fields $\varphi^\pm$ are commuting and the fields $ b, e$ are anticommuting elements of the Grassman algebra. The fields $\varphi^\pm, b, e$ belong to the adjoint representation of the group $SU(2)$.

We firstly consider the topologically trivial sector. Shifting the fields $\varphi^-$ along the third axis
 \begin{equation}
 \varphi_i^-\rightarrow  \varphi_i^--\hat{\mu}; \quad \hat{\mu}=\delta^{i3}mg^{-1}
\label{9}
\end{equation}
we obtain the following Lagrangian
\begin{equation}
L=-\frac{1}{4}F^i_{\mu \nu}F^i_{\mu \nu}+(D_\mu \varphi^+)^i(D_\mu \varphi^-)^i+i(D_\mu b)^i(D_\mu e)^i- (D_\mu \varphi^+)^i(D_\mu \hat{\mu})^i
\label{10}
\end{equation}
This Lagrangian is obviously invariant with respect to the "shifted" gauge transformations
\begin{eqnarray}
\delta A_\mu^a=\partial_\mu\eta^a+g\varepsilon^{abc} A_\mu^b\eta^c\nonumber\\
\delta\varphi^+_a=g\varepsilon^{abc}\varphi^+_b\eta^c\nonumber\\
\delta\varphi^-_a=-m\varepsilon^{a3c} \eta^c +g\varepsilon^{abc}\varphi^-_b\eta^c\nonumber\\
\delta b^a=g\varepsilon^{abc}b^b\eta^c\nonumber\\
\delta e^a=g\varepsilon^{abc}e^b\eta^c\nonumber\\
i,j=1,2,3; \quad a,b=1,2.
\label{11}
\end{eqnarray}
The Lagrangian (\ref{10}) is also invariant with respect to the transformation of supersymmetry
 \begin{equation}
 \delta\varphi^-_i=i b_i \epsilon; \quad \delta e_i=\varphi^+_i\epsilon; \quad \delta b^i=\delta\varphi^+_i=0
\label{12}
\end{equation}
This is a new symmetry which was absent in the standard Yang-Mills Lagrangian and which plays the main role in the proof of decoupling of unphysical fields $\varphi^\pm, b, e$ from the physical components of $A_\mu^{tr}$.

Note that the transformations shift the fields $\varphi^-_{1,2}$ by the arbitrary functions. Therefore as in the Higgs model except for the gauge field $A_\mu$ the new gauge field $\varphi^-_{1,2}$ arises. Let us choose the gauge 
 \begin{equation}
\varphi^-_{1,2}=0
\label{13}
\end{equation}
This gauge is obviously algebraic and does not require the introduction of the Faddeev-Popov ghosts. At the same time the condition (\ref{13}) is manifestly Lorentz invariant. Therefore in this model  we succeeded to introduce the algebraic gauge which is manifestly Lorentz invariant, and as we shall see preserving the renormalizability of the theory. The remaining gauge invariance, related to the rotations around the third axis in the charge space is Abelian and does not produce the Gribov ambiguity. 

However the gauge $\varphi^-_{1,2}=0$ is still not unique. Applying to the fields $\varphi^-_{1,2}$ the gauge transformations (\ref{11}) we get the equations which should be satisfied by the gauge function to provide the uniqueness of the gauge
\begin{equation}
(m-g\varphi^-_3)\eta^2=0; \quad (m-g\varphi^-_3)\eta^1=0;
\label{14}
\end{equation}
Note that in perturbation theory, when $\eta^a= \eta^a_0+g\eta^a_1+ \ldots$ the only solution of the equations (\ref{14}) is $\eta^{1,2}=0$. But for the values of $\varphi^-_3$, for which $(m-g\varphi^-_3)=0$, the choice of the gauge as before is not unique.

To get rid of this non uniqueness we shall make the change of variables in the classical Lagrangian
\begin{eqnarray}
\varphi^-_3=-\frac{m}{g}(\exp\{\frac{gh}{m}\}-1); \quad \varphi^-_{1,2}= M\tilde{\varphi}^-_{1,2}; \quad  \varphi^+_{1,2}=M^{-1}\tilde{\varphi}^+_{1,2}\nonumber\\
\varphi^+_3=M^{-1}\tilde{\varphi}^+_3; \quad e=M^{-1}\tilde{e}; \quad  b=M\tilde{b},
\label{15}
\end{eqnarray}
Instead of the gauge $\varphi^-_{1,2}=0$ we impose the condition
\begin{equation}
\tilde{\varphi}^-_{1,2}=0.
\label{17}
\end{equation}
In this way we get
\begin{equation}
\delta\tilde{\varphi}^-_{1,2}=\pm m \eta_{1,2}
\label{18}
\end{equation}
As one can see the non uniqueness of the gauge fixing is completely absent. The effective Lagrangian in the gauge (\ref{18})  looks as follows 
\begin{eqnarray}
L_{ef}=-\frac{1}{4}F^i_{\mu \nu}F^i_{\mu \nu}+\partial_\mu h\partial_\mu\tilde{\varphi}^+_3-\frac{g}{m}\partial_\mu h\partial_\mu h\tilde{\varphi}^+_3\nonumber\\
-[(D_\mu\tilde{b})+\frac{g}{m}\tilde{b}\partial_\mu h]^i[(D_\mu\tilde{e})-\frac{g}{m}\tilde{e}\partial_\mu h]^i\nonumber\\
+mg(A_\mu)^2\tilde{\varphi}^+_3+g\partial_\mu h A_\mu^a\tilde{\varphi}^+_a+m\varepsilon^{3ab}\tilde{\varphi}^+_a\partial_\mu A_\mu^b
\label{19}
\end{eqnarray}
The free propagators determined by the Lagrangian (\ref{19}) are
\begin{eqnarray}
\triangle(A_\mu^i A_\nu^j)=-i\delta^{ij}\frac{T_{\mu\nu}}{p^2}; \quad \triangle(A^a_\mu \tilde{\varphi}^+_b)=\varepsilon^{3ab}\frac{p_\mu}{mp^2}\nonumber\\
\triangle(\tilde{b}^i\tilde{e}^j)=\frac{\delta^{ij}i}{p^2}; \quad \triangle(h \tilde{\varphi}^+_3)=\frac{1}{p^2}
\label{20}
\end{eqnarray}
where $T_{\mu\nu}$ is the transversal projector. The remaining propagators  corresponding to the Abelian subgroup of rotations around the third axis, one can put equal to
\begin{equation}
\triangle(A_\mu^3A_\nu^3)=-i\frac{T_{\mu\nu}}{p^2}; \quad \triangle(A_\mu^3\tilde{\varphi}^+_3)=-\frac{p_\mu}{mp^2}
\label{21}
\end{equation}
It is easy to calculate the divergency index of arbitrary diagram. It is equal to
\begin{equation}
n=4-2L_{\tilde{\varphi}^+_3}-2L_{\tilde{\varphi}^+_a}-L_A-L_e-L_b-L_h
\label{22}
\end{equation}
where $L_c$ denotes the number of external lines of the field $c$. As the interaction Lagrangian includes only the trilinear vertices with one derivative and four linear vertices without derivatives the theory is obviously renormalizable.

\section{Unitarity of the theory in the physical space.}

Spectrum of our theory includes several unphysical particles. They are zero and the third components of $A^i_\mu$,  the fields $\varphi^\pm_a$, and anticommuting fields 
$b^a, e^a$. Let us remind that the original Lagrangian was invariant with respect to the gauge transformations (\ref{11})  and transformations of supersymmetry (\ref{12}). In terms of new variables the asymptotic gauge transformations do not change, and the supersymmetry transformations are changed as follows
\begin{eqnarray}
\delta\tilde{\varphi}^-_a=i\tilde{b}^a \epsilon\nonumber\\
\delta h=i\tilde{b}^3  \epsilon\nonumber\\
\delta\tilde{e}^i=\tilde{\varphi}^+_i
\label{23}
\end{eqnarray}
The remaining components of asymptotic fields are not transformed. We say nothing about the Abelian gauge transformations, related to rotations around the third axis. These transformations do not introduce any complications. 

Let us fix the gauge, adding to the effective action the following expression
\begin{equation}
s_1\int d^4x \bar{c}^a\tilde{\varphi}^-_a= \int d^4x[\lambda^a\tilde{\varphi}^-_a+\bar{c}^a(c^a-\tilde{b}^a)]
\label{24}
\end{equation}
where $s_1$ is the nilpotent operator, similar to the usual BRST-operator, determined by the gauge transformation, leaving the effective Lagrangian (\ref{19}), written in terms of transformed variables, invariant and the action of operator $s_1$ on the ghost fields and the field $\lambda$ is defined by the formula
\begin{equation}
(s_1c)^a=0; \quad (s_1\bar{c})^a=\lambda^a; \quad (s_1\lambda)^a=0.
\label{25}
\end{equation}
The effective action with the fixed gauge looks as follows
\begin{equation}
A_{ef}=\int d^4x(L(x)+\lambda^a\tilde{\varphi}^-_a+m\bar{c}^ac^a-\bar{c}^a\tilde{b^a})
\label{26}
\end{equation}
where $L$ denotes the Lagrangian (\ref{19}), invariant with respect to simultaneous gauge transformations and transformations of supersymmetry, written in terms of variables $\lambda, h, \tilde{\varphi}^\pm, A_\mu, \tilde\varphi^+_3, \tilde{b},  \tilde{e}$.
The canonical gauge fixing does not include the term $\bar{c}^a\tilde{b}^a$, but this term may be easily generated be the change of variables $c \rightarrow c-\tilde{b}m^{-1}$. Performing explicitly the integration over $\bar{c}^a, c^a$ we get the action, which is invariant under the simultaneous BRST-transformations and transformations of supersymmetry with the change  $c \rightarrow \tilde{b}m^{-1}$.

According to the Noether theorem the invariance of the effective action leads to existence of the conserved charge $Q$, which allows to separate physical excitations by requiring their annihilation by the asymptotic charge $Q^0$
\begin{equation}
Q^0|\Phi>_{as}^{ph}=0
\label{27}
\end{equation}   
where $Q^0$ is the asymptotic operator acting on the asymptotic fields as follows
\begin{eqnarray}
Q^0A^a_\mu=i\frac{\partial_\mu\tilde{b}^a\epsilon}{m}; \quad Q^0\tilde{b}^a=0;\nonumber\\
Q^0 h=i\tilde{b}^3\epsilon; \quad Q^0\tilde{b}^3=0;\nonumber\\
Q^0\tilde{e}^a=\tilde{\varphi}^+_a\epsilon; \quad Q^0\tilde{\varphi}^+_a=0;\nonumber\\
Q^0\tilde{e}^3=\tilde{\varphi}^+_3\epsilon; \quad Q^0\tilde{\varphi}^+_3=0.
\label{28}
\end{eqnarray}
It may be seen from the preceding formula that unphysical fields enter in the form of BRST-doublets. If we identify the field $\tilde{e}^a$ with the anti ghost field $\bar{c}$, and the field $\tilde{b}^am^{-1}$ with ghost field $c$, these transformations provide the decoupling of the fields $\tilde{\varphi}^+_a, \tilde{e}^a, \tilde{b}^a$ and unphysical components of the Yang-Mills field from the transversal components. The remaining transformations provide the decoupling of the components $\tilde{e}^3, \tilde{b}^3, h, \tilde{\varphi}^+_3$.
Fixing in addition the Abelian degree of freedom of the vector field, we conclude, that all unphysical fields are decoupled from the transversal ones.

Till now we considered the Yang-Mills fields with zero mass. This theory is the basis of Quantum Chromodynamics (QCD). The modern QCD uses one more essential postulate, the hypothesis about confinement of color. The confinement of color objects can not be explained in the framework of perturbation theory with respect to the coupling constant. For explanation of the color confinement lattice simulations are commonly used. The models used for this purpose as a rule contain the quasi particle excitations, which we call solitons. On the other hand it is known that classical equation of motions in the standard formulation of the Yang-Mills theory have no soliton solutions
 (\cite{Co}, \cite{De}, \cite{Pa}). However the arguments forbidding the existence of solitons in the Yang-Mills theory are not applicable to the modified formulation presented above. In the next sections we shall show that classical solutions of the soliton type indeed exist in QCD based on the modified formulation of the Yang-Mills theory.

\section{Expansion over the coupling constant.}

The original Lagrangian we choose in the form
\begin{equation}
L=-\frac{1}{4}F_{\mu\nu}^iF_{\mu\nu}^i+\frac{1}{2}(D_\mu \varphi)^i(D_\mu \varphi)^i -\frac{1}{2}(D_\mu \chi)^i(D_\mu \chi)^i+i(D_\mu b)^i(D_\mu e)^i
\label{29}
\end{equation}
as in the previous sections we consider for simplicity the group $SU(2)$. In this formula we introduced explicitly instead of the fields $\varphi^\pm$ the fields $\chi$, which have the negative energy. The fields $\varphi,\chi$, are commuting and the fields $b,e$ anticommute. 
All these fields belong to the adjoint representation of the group $SU(2)$.

We consider the fields $\varphi,\chi$ with nontrivial asymptotic behavior:
\begin{equation}
|\varphi|\rightarrow|\frac{m}{g}|; \quad |\chi|\rightarrow|\frac{m\alpha}{g}|; \quad |\alpha|\leq 1; \quad r=|\textbf{x}|; \quad r\rightarrow \infty
\label{30}
\end{equation}
Parameter $\alpha \rightarrow 1$ when $g \rightarrow 0$. For example
\begin{equation}
\alpha=\frac{g^{-n}-g^n}{g^{-n}+g^n}=1-g^{2n}+ \ldots
\label{31}
\end{equation}
so that $\alpha=1-O(g^{2n})$. Choosing $n$  sufficiently big we shall get in the formal perturbation theory the results coinciding within the standard Yang-Mills theory to the arbitrary order in $g$. In equation (\ref{30})
$m$ is a constant having the dimension of mass. 

We call the formal perturbation theory the power series over the coupling constant, independently of the fact is it convergent or divergent. Even the separate terms in this series may not exist in the limit when some intermediate regularization is removed. In QCD the limit for the separate terms of the series may not exist due to infrared divergencies. Numerous attempts to constract the Yang-Mills theory analogously to quantum electrodynamics failed due  to nonlinear interaction of the Yang-Mills quanta.

If the coupling constant is small as in electrodynamics the usual relation of the type of unitarity or causality conditions are fulfilled in the formal perturbation theory at any order in the coupling constant.  However in QCD the coupling constant is not small and even the separate terms in the formal perturbation theory may not exist due to infrared singularities. Nevertheless usually one insists on the validity in the formal perturbation theory for the Yang-Mills field of the conditions of unitarity and causality. This point of view is supported by the observation of jets, which is the evidence of foundation of QCD on some nonabelian gauge theory. However no rigorous statements can be done, as hadronization process which plays the important role in the formation of jets is essentially nonperturbative.

The only sensible object in perturbative quantum theory of the Yang-Mills field  may be the generating functional for the gauge invariant operators. Below we shall show that in our formulation this functional coincides with the standard expression.

Firstly we consider the topologically trivial sector, corresponding to the perturbation theory. In this case we can choose the direction in which the fields do not vanish as the third axis in the charge space. Making the shift of variables $\varphi,\chi$
\begin{equation}
\varphi^i=\tilde{\varphi}^i+\delta^{i3}mg^{-1}; \quad \chi^i=\tilde{\chi}^i-\delta^{i3}m\alpha g^{-1}
\label{32}
\end{equation}
we get the Lagrangian in which the fields $\tilde{\varphi}, \tilde{\chi}=0$ at infinity which is necessary for the construction of perturbation theory.
We want to prove that the scattering matrix, obtained after the shift, in the framework of perturbation theory (in the limit $\alpha\rightarrow 1$) coincides with the standard scattering matrix in the Yang-Mills theory.
If $\alpha \neq 1$, one can speak about the coincidence of the scattering matrices up to arbitrary order of formal perturbation theory.
Of course the on shell scattering matrix strictly speaking does not exists doe to infrared singularities, nevertheless it is possible to speak about nullification of the matrix elements, corresponding to transitions between the states which contain only physical excitations and the states, containing some unphysical excitations. In both cases the physical excitations correspond to the transversal components of the Yang-Mills field. 

In the topologically nontrivial sectors our theory differs of the standard: the Yang-Mills theory in the standard formulation has no soliton excitations. At the same time the modified formulation describes the classical solitons.  

At present we are interested only in the results of perturbation theory, therefore we can put the parameter $\alpha=1$, as $\alpha=1-O(g^{2n})$, where $n$ is an arbitrary number.
Clearly in this case no mass term for the Yang-Mills field arises, as due to the different signs before the terms depending on the fields $\varphi$ and  $\chi$
their contributions to the mass term for the Yang-Mills field are mutually compensated.

The Lagrangian describing the modified theory after the shift
(\ref{32}) looks as follows
\begin{eqnarray}
 L=-\frac{1}{4}F_{\mu\nu}^iF_{\mu\nu}^i+ D_\mu \tilde{\varphi}_+^iD_\mu \tilde{\varphi}_-^i+iD_\mu \tilde{b}^i D_\mu \tilde{e}^i+\nonumber\\
  m\frac{1+\alpha}{\sqrt{2}}D_\mu
  \tilde{\varphi}_+^i\varepsilon^{ij3}A_\mu^j+m\frac{1-\alpha}{\sqrt{2}}
D_\mu \tilde{\varphi}_-^i\varepsilon^{ij3}A_\mu^j +
\frac{m^2(1-\alpha^2)}{2}A_\mu^a A_\mu^a
 \label{33}
 \end{eqnarray}
Here the obvious notations are used
\begin{equation}
\tilde{\varphi}_\pm^i=\frac{\tilde{\varphi}^i \pm \tilde{\chi}^i}{\sqrt{2}}
\label{34}
\end{equation}
Here $i,j=1,2,3,; a=1,2.$

This Lagrangian for any $\alpha$ is invariant with respect to the "shifted" gauge transformations
\begin{eqnarray}
\delta A_\mu^i=\partial_\mu \eta^i+g\varepsilon^{ijk}A_\mu^j \eta^k \nonumber\\
\delta \tilde{\varphi}_-^1=-\frac{1+\alpha}{\sqrt{2}}m\eta^2+g\varepsilon^{1jk}\tilde{\varphi}_-^j\eta^k \nonumber\\
\delta \tilde{\varphi}_+^1=-\frac{1-\alpha}{\sqrt{2}}m\eta^2+g\varepsilon^{1jk}\tilde{\varphi}_+^j\eta^k \nonumber\\
\delta \tilde{\varphi}_-^2=\frac{1+\alpha}{\sqrt{2}}m\eta^1+g\varepsilon^{2jk}\tilde{\varphi}_-^j\eta^k \nonumber\\
\delta \tilde{\varphi}_+^2=\frac{1-\alpha}{\sqrt{2}}m\eta^1+g\varepsilon^{2jk}\tilde{\varphi}_+^j\eta^k \nonumber\\
\delta \tilde{\varphi}_-^3=g\varepsilon^{3jk}\tilde{\varphi}_-^j\eta^k \nonumber\\
\delta \tilde{\varphi}_+^3=g\varepsilon^{3jk}\tilde{\varphi}_+^j\eta^k \nonumber\\
\delta \tilde{b}^i=g\varepsilon^{ijk}\tilde{b}^j\eta^k \nonumber\\
\delta \tilde{e}^i=g\varepsilon^{ijk}\tilde{e}^j\eta^k
\label{35}
\end{eqnarray}

In perturbation theory the Gribov ambiguity is absent, so we can choose a gauge
$\partial_\mu A_\mu=0$, introducing simultaneously Faddeev-Popov ghosts $\bar{c}, c$.

The scattering matrix at $\alpha=1$ may be written as a path integral
\begin{eqnarray}
  S=  \int d\mu \exp\{i[\int d^4x (-\frac{1}{4}F_{\mu\nu}^iF_{\mu\nu}^i+ D_\mu \tilde{\varphi}_+^iD_\mu \tilde{\varphi}_-^i+\nonumber\\
 +\lambda^i\partial_\mu A_\mu^i+i\partial_\mu\bar{c^i}D_\mu c^i+ +iD_\mu \tilde{b}^iD_\mu \tilde{e}^i +m\sqrt{2}D_\mu \tilde{\varphi}_+^i\varepsilon^{ij3}A_\mu^j)]\}
\label{36}
\end{eqnarray}
where the integration measure $d\mu$ is the product of differentials of all the fields entering the Lagrangian.

For $\alpha=1$ the Lagrangian (\ref{33}) is invariant with respect to the supersymmetry transformation
\begin{equation}
 \delta \tilde{\varphi}_-^i=\tilde{b}^i \epsilon; \quad \delta \tilde{e}^i=\tilde{\varphi}_+^i\epsilon; \quad \delta\tilde{b}^i=\delta \tilde{\varphi}_+^i=0.
 \label{37}
 \end{equation}
It is easy to see that these transformations are nilpotent.
\begin{equation}
\delta^2\tilde{\varphi}_-^i=0; \quad \delta^2\tilde{e}^i=0
\label{38}
\end{equation}
This invariance provides the decoupling of excitations corresponding to the fields
$\tilde{\varphi}_\pm, \tilde{b}, \tilde{e}$.
As in the previous section the invariance of the effective action with respect to the BRST-transformations and the supersymmetry transformations according to the Noether theorem generates the conserved charges $Q_B, Q_S$, and the asymptotic states may be chosen in such a way, that they satisfy the equations
\begin{equation}
Q_B^0|\psi>_{ph}=0; \quad Q_S^0|\psi>_{ph}=0; \quad [Q^0_B,Q^0_S]_+=0
\label{39}
\end{equation}
where $Q_B^0$ and $Q_S^0$ are asymptotic charges.
Any vector which satisfies the equations (\ref{39}) has a form
\begin{equation}
|\psi>_{ph}=|\psi>_{tr}+|N>
\label{40}
\end{equation}
where $|\psi>_{tr}$ is a vector, which contains only  transversal quanta of the Yang-Mills  field, and  $|N>$ is a vector with zero norm.
From this result it follows that the scattering matrix in our formulation coincides with the scattering matrix in theYang-Mills theory.

The proof however was formal, as the scattering matrix does not exist in the Yang-Mills theory due to infrared singularities. We can however speak about nullification of the matrix elements corresponding to the transitions between physical and unphysical states.

The only nontrivial objects making sense in the perturbative Yang-Mills theory are the correlation functions of gauge invariant operators. It is easy to see that these correlation functions coincide in the standard and modified formulation up to arbitrary order of perturbation theory. Indeed, one can repeat the considerations presented above and show that these correlation functions are given by the path integral
\begin{eqnarray}
 Z=\int d\mu\{\exp[i\int dx(-\frac{1}{4}F_{\mu\nu}^iF_{\mu\nu}^i+ D_\mu \tilde{\varphi}_+^iD_\mu \tilde{\varphi}_-^i+m\sqrt{2}D_\mu \tilde{\varphi}_+^i\varepsilon^{ij3}A_\mu^j\nonumber\\+\lambda^i\partial_\mu A_\mu^i+i\partial_\mu \bar{c}^iD_\mu c^i+iD_\mu b^iD_\mu e^i+J(x)O(x))]\}
\label{41}
\end{eqnarray}
where  $O(x)$ is some gauge invariant operator depending only on $A_\mu(x)$, and $J(x)$ is the source.
The boundary conditions for all the fields in the equation (\ref{41}) correspond to the vacuum states.
In the generating functional(\ref{41}) for $m=0$ we can integrate explicitly over the fields
$\tilde{\varphi}_\pm, e, b$. The determinants which arise after such integration compensate each other, as the fields $\tilde{\varphi}_\pm$ and $b,e$ obey different statistics, and we obtain for the generating functional of the correlation functions of gauge invariant operators in the Yang-Mills theory the expression
\begin{equation}
Z=\int d\tilde{\mu}\{\exp[i\int dx(-\frac{1}{4}F_{\mu\nu}^iF_{\mu\nu}^i +\lambda^i\partial_\mu A^i_\mu+i\partial_\mu\bar{c}^iD_\mu c^i+J(x)O(x))]\}
\label{42}
\end{equation}
where the integration measure $d\tilde{\mu}$ is the product of differentials
 \begin{equation}
 d\tilde{\mu}=dA_\mu^id\lambda^jd\bar{c}^k dc^l
 \label{43}
 \end{equation}
We would arrive to the same conclusion if we worked in the gauge applicable beyond perturbation theory, for example
$\tilde{\varphi}_-^a=0, a=1,2; \quad \partial_\mu A_\mu^3=0$. Starting from this gauge we can pass to any admissible gauge. In the next section we consider the classical theory in the Hamiltonian gauge
 $A_0=0$.

\section{ Soliton excitations in the modified Yang-Mills theory.}

In this section we shall show that the model, constructed above, has nontrivial soliton excitations of the t'Hooft-Polyakov magnetic monopole type. 

We consider the classical solutions for the action (\ref{29}) and look for the classical solitons, which have for large
 $r$ the asymptotics
\begin{equation}
\varphi^i\rightarrow\frac{x^im}{rg}; \quad \chi^i\rightarrow-\frac{x^im\alpha}{rg};
\label{44}
\end{equation}
We are dealing with the stationary solutions in the gauge $A_0=0$.

We shall work in the topologically nontrivial sector and look for the nonperturbative soliton solutions of the classical equations of motion.
\begin{eqnarray}
D_iF^a_{ij}+g\varepsilon^{alm}(D_j\varphi)^l \varphi^m-g\varepsilon^{akn}(D_j \chi)^k\chi^n=0; \quad A^a_i\rightarrow \varepsilon^{aij}\frac{x^j}{gr^2}, r\rightarrow \infty \nonumber\\
D_i(D_i \varphi)^n=0; \quad \varphi^n(x)\rightarrow\frac{x^nm}{gr}, r\rightarrow \infty\nonumber\\
D_i(D_i\chi)^n=0; \quad \chi^n(x)\rightarrow -\frac{\alpha x^nm}{gr}, r\rightarrow \infty.
\label{45}
\end{eqnarray}
The asymptotic conditions we choose provide fast decreasing of the covariant derivatives of the fields
 $\varphi, \chi$, which is important for the finiteness of the soliton energy. Having in mind the same goal we consider only solutions which are non singular at $r\rightarrow 0$. Now we cannot neglect the terms which are small in the formal perturbation theory, as we are looking for soliton solutions which cannot be obtained in perturbation theory. Note also that in QCD the coupling constant is not small, therefore the conclusions done in the formal perturbation theory may be wrong.

We shall use the t'Hooft-Polyakov ansatz
\begin{eqnarray}
A^i_a(x)=\varepsilon^{aij}\frac{x^j}{r}W(r); \quad \varphi^i(x)=\delta^{ai}\frac{x_a}{r}F(r) \nonumber \\
\chi^i(x)=\delta^{ai}\frac{x_a}{r}G(r); \quad A_0(x)=0, \nonumber\\
r\rightarrow \infty, W(r)\rightarrow (gr)^{-1}, F(r)\rightarrow F\cosh{\gamma}, G(r)\rightarrow F\sinh{\gamma},\nonumber\\ F\cosh{\gamma}=\frac{m}{g}; \quad F\sinh{\gamma}=-\alpha \frac{m}{g}.
\label{46}
\end{eqnarray}
If $g$ is small, $\alpha\rightarrow 1$, as it happens in the electro-weak models based on the mechanism of Brout-Englert-Higgs, then  $\varphi(x)\simeq \chi(x)$
and the equation for the Yang-Mills field has the same form as in the standard theory of Yang-Mills.  This equation has no soliton solutions.
However we can also consider the theories in which the constant
 $g$ is not small (for example QCD).

The equations(\ref{45}) may be rewritten in terms of functions
\begin{equation}
K(r)=1-grW(r); \quad J(r)=F(r)rg; \quad Y(r)=G(r)rg
\label{47}
\end{equation}
\begin{eqnarray}
r^2\frac{d^2K}{dr^2}=(K^2+J^2-Y^2-1)K(r); \quad K(r)\rightarrow 0, r\rightarrow \infty\nonumber\\
r^2\frac{d^2J}{dr^2}=2K^2J; \quad J(r)\rightarrow Frg\cosh{\gamma};  r\rightarrow \infty \nonumber\\
r^2\frac{d^2Y}{dr^2}=2K^2Y; \quad Y(r)\rightarrow Frg\sinh{\gamma}=-\alpha Frg\cosh{\gamma};  r\rightarrow \infty
\label{48}
\end{eqnarray}
Following the paper (\cite{JZ}) we choose the following ansatz for solutions
\begin{eqnarray}
J(r)=\Lambda(r)\cosh{\gamma}; \quad Y(r)=\Lambda(r)\sinh{\gamma};\nonumber\\
\Lambda(r)\cosh{\gamma}\rightarrow Frg\cosh{\gamma}; \quad \Lambda(r)\sinh{\gamma}\rightarrow Frg\sinh{\gamma}.
\label{49}
\end{eqnarray}

After that the equations (\ref{48}) acquire the form
\begin{eqnarray}
r^2\frac{d^2K}{dr^2}=(K^2+\Lambda^2-1)K; \quad K\rightarrow 0, r\rightarrow \infty, \nonumber\\
r^2\frac{d^2\Lambda}{dr^2}=2K^2\Lambda; \quad \Lambda(r)\rightarrow Frg; \quad r\rightarrow \infty.
\label{50}
\end{eqnarray}
The solutions of this system are well known
(\cite{PZ},\cite{Bog})
\begin{equation}
K(r)=\frac{rgF}{\sinh{rgF}}; \quad \Lambda(r)=\frac{rgF}{\tanh{grF}}-1.
\label{51}
\end{equation}

It is not difficult to calculate the energy corresponding to this solution. Obviously this energy is positive and limited. It is equal to the energy of magnetic monopole
\begin{eqnarray}
E=\int d^3x[\frac{1}{4}F_{lm}^iF_{lm}^i+\frac{1}{2}(D_l\varphi)^i(D_l\varphi)^i -\frac{1}{2}(D_l\chi)^i(D_l\chi)^i]=\nonumber\\
\int d^3x[\frac{1}{4}F_{lm}^iF_{lm}^i+\frac{1}{2}(D_l\Lambda)^a(D_l\Lambda^a)]
\label{52}
\end{eqnarray}

One can calculate also the magnetic field created by this solution. Using the gauge invariant definition of the electromagnetic stress tensor, we get 
\begin{equation}
F_{\mu\nu}=\hat{\Lambda}^aF_{\mu\nu}^a-g^{-1}\varepsilon^{abc}\hat{\Lambda}^a(D_\mu\hat{\Lambda})^b(D_\nu\hat{\Lambda})^c
\label{53}
\end{equation}
where $\hat{\Lambda}^a=\frac{\Lambda^a}{|\Lambda|}; \quad
|\Lambda|=(\sum_a\Lambda^a\Lambda^a)^{1/2}$ we find that the excitation we consider is the magnetic monopole, creating the  magnetic field
\begin{equation}
B^i(x)=\frac{x^i}{gr^3}
\label{54}
\end{equation}
One can see that even for large $g$ the mass and the magnetic field of the monopole do not depend on $\gamma$, and are determined by the constants $F$ and $g$.

The presentation in this section follows the paper (\cite{Sl4}).

\section{Results.}
 
In this review we demonstrated that many facts which we considered as firmly established in the theory of Yang-Mills (impossibility to quantize the theory beyond perturbation theory, unavoidable breaking by algebraic renormalisable gauge of the manifest Lorentz invariance of the theory, the absence of classical solutions with the finite limited energy) in fact are related to the specific formulation of the theory. It is possible to give the alternative
formulation of the theory which giving in the formal perturbation theory the same results that the standard one, which allows to overcome these difficulties. This formulation follows the general tendency of development of the gauge fields, introduction of new unphysical degrees of freedom and enlarging the group of the symmetry of the theory. In this way it is possible in particular to quantize the Yang-Mills theory beyond perturbation theory. Naturally it does not solve the problem of calculations beyond perturbation theory, however it shows that the absence of soliton excitations which is thought necessary to provide the confinement of color, is not the unavoidable feature of the theory.  As was shown this theory allows the alternative formulation which leads to the existence of the soliton solutions of the classical equations. This formulation also allows to construct the infrared regularization,which preserves the manifest gauge and Lorentz invariance (\cite{Sl5}).

\section{Acknowlegements.}

I wish to thank for the useful discussions L.Alwarez-Gome, D.V.Bykov, L.D.Faddeev, V.A.Rubakov, and R.Stora. This work was done at Steklov Mathematical Institute and supported by Russian Scientific Foundation (project 140500005)
 \end{document}